\documentstyle[graphicx]{mn}

\newif\ifAMStwofonts
\AMStwofontstrue

\def\2M{{2MASS 0918+2117}}

\def\xmm{{\it XMM-Newton}}
\def\chandra{{\it Chandra}}

\def\et{{et al.\ }}

\newcommand{\ls}{\mathrel{\hbox{\rlap{\hbox{\lower4pt\hbox{$\sim$}}}\hbox{$<$}}}}
\newcommand{\gs}{\mathrel{\hbox{\rlap{\hbox{\lower4pt\hbox{$\sim$}}}\hbox{$>$}}}}

\def\arcs{{\hbox{$^{\prime\prime}$}}}

\def\H0{{\rm ~km~s^{-1}~Mpc^{-1}}}

\def\et{{et al.}}

\title[\xmm\ observation of \2M]
        {Comparison of high and low state X-ray spectra in the Type 1.5 QSO \2M}
\author[K.A.Pounds \et]
        {K.A.Pounds$^{1}$,
        and B.J.Wilkes$^{2}$\\
$^1$ Department of Physics and Astronomy, University of Leicester,
Leicester, LE1 7RH, UK\\
$^2$ Harvard/Smithsonian Center for Astrophysics, Cambridge, MA 02138, USA\\}

\date{Accepted ; Submitted ; Revised }
\pagerange{\pageref{firstpage}--\pageref{lastpage}}
\pubyear{2007}
\begin{document}
\maketitle
\label{firstpage}

\begin{abstract} 
When observed by \xmm\ in 2003 the type 1.5 QSO \2M\ was found to be in a low state, with an X-ray flux $\sim$4-5 times fainter than
during an earlier \chandra\ observation.  The 2-6 keV spectrum was unusually hard (photon index $\Gamma$$\sim$1.25), with evidence for a reflection-dominated
continuum, while a soft excess visible below $\sim$1 keV prevented confirmation of the anticipated low energy absorber (Wilkes \et\ 2005). 
In a second \xmm\ observation in 2005 the
X-ray flux is found to have recovered, with a 2-10 keV continuum spectrum now typical of a broad-line active galaxy ($\Gamma$$\sim$2) and 
a deficit of flux below $\sim$1 keV indicative of
continuum absorption in a column N$_{H}$$\sim$$4\times 10^{21}$ cm$^{-2}$. We find the preferred ionisation state of the absorbing
gas to be low, which then leaves a residual soft excess of similar spectral form and flux to that found in the 2003 \xmm\
observation. Although observed at different epochs we note that dust in the absorbing column could also explain the red nucleus and 
strong optical polarisation  of \2M.
\end{abstract}

\begin{keywords} galaxies: active -- galaxies:general -- galaxies: individual:2MASS 0918+2117 -- galaxies:QSO -- X-ray:galaxies 
\end{keywords}

\section{Introduction}
The Two Micron All-Sky Survey (2MASS) has revealed many highly reddened active galaxies (AGN) whose number density rivals that of
optically selected AGN. Spectroscopic follow-up of red candidates showed $\sim$75 percent to be previously unidentified
emission-line AGN, with $\sim$80 percent of those showing the broad optical emission lines of Type 1 Seyfert galaxies and QSOs
(Cutri \et\  2001). These objects often have unusually high optical polarization levels, with $\sim$10 percent showing $P>3$ percent
indicating a significant contribution from scattered light (Smith \et\ 2002) and suggesting substantial obscuration toward the
nuclear energy source. \chandra\ observations of a sample of 2MASS AGN found them to be X-ray weak with generally flat (hard)
spectra (Wilkes \et\ 2002).

While cold absorption is an expected signature in reddened AGN, \xmm\ follow-up observations of a subset of five 2MASS AGN (Wilkes
\et\ 2005; hereafter W05) showed a mixed picture. The longer \xmm\ exposures confirmed a substantial column of cold matter
(N$_{H}$$\sim$$10^{22}$ cm$^{-2}$) in 2 cases, both optical type 2 AGN. In 2 further cases, both optical type 1, low energy
absorption was apparent in one, although complicated by  a soft emission component (Pounds \et\ 2005). The 5th
object in the \xmm\ sample, \2M, was potentially the most interesting, being optically classified as an intermediate Type 1.5 QSO at a 
redshift of z = 0.149 (Cutri \et\ 2003).
However the data on \2M\ were of low quality due to the source being unexpectedly faint.

\2M\ was selected for the initial \chandra\ sample due to its unusually `red' nucleus (J-K$_{s}$$\sim$2.23) and high optical polarisation
of $\sim$6.4\% (Smith et al 2002, 2003). Given that evidence of strong nuclear obscuration, the \chandra\ detection of a `normal'
type 1 X-ray spectrum ($\Gamma$$\sim$1.9), with only marginal evidence of a cold absorber, was a
surprise (W05). That issue remained unclear after the the first \xmm\ observation of \2M\ in 2003, which found a much 
harder overall spectrum ($\Gamma$$\sim$1.25) than seen by \chandra, while the X-ray flux was a factor $\sim$4--5 fainter. 

Extrapolating the hard (2-10 keV) power law down to 0.3 keV (figure 1 in W05) revealed a clear soft excess, the presence of which
potentially could have `hidden' the effects of low energy absorption in the \chandra\ spectrum, while an excess of counts in the
highest energy channels was suggestive of reflection. Including a strong cold reflection component in the model, W05 found the power
law index increased by $\sim$0.3. Although not required by the data, inclusion of an absorbing column at the 1 $\sigma$ upper bound
indicated by \chandra\ further increased the intrinsic continuum slope closer to the `normal' $\Gamma$$\sim$1.8-2 range for unobscured
AGN (Nandra and Pounds 1994). 

As the soft excess remained, a broad gaussian emission `line' was included in the final 0.3-10 keV spectral model.  The parameters of the
best-fit model were an underlying power law index $\Gamma$=1.65$\pm$0.33, with cold reflection R=6$\pm$3, and a gaussian emission 
line at 0.44$\pm$0.13 keV, with line width $\sigma$=75 eV and flux $\sim$$4\times 10^{-5}$ photon cm$^{-2}$s$^{-1}$.

In summary, the first \xmm\ observation of \2M\ found the source in a low flux state, with a hard power law continuum probably
dominated by a strong reflection component (W05). The hard continuum, in turn, allowed the soft excess to stand out, but 
evidence of the X-ray absorption that might be expected for such a highly reddened AGN remained unclear.

Here we report the outcome of a second, longer \xmm\ observation of \2M, finding the X-ray flux to be now a factor $\sim$2 brighter
than when observed by \chandra. 

\section{Observation and data reduction} 
The new \xmm\ observation of \2M\ took place on 2005 November 15/16. The EPIC pn (Str\"{u}der \et 2001) and MOS1 and MOS2 (Turner \et\
2001) cameras obtained CCD-resolution spectra over the energy band $\sim$0.2--10 keV. Each EPIC camera was set in large window mode, with
the thin filter (as in the first observation). Unfortunately the source was too weak to get useful high resolution spectra from the grating spectrometers,
while the only active OM channel (UVW1) yielded a magnitude of 19.3, identical with that in the 2003 observation. We conclude from the latter that the 
AGN is
probably highly obscured in this UV band. 

EPIC X-ray data were screened with the XMM SAS v7.0 software and events corresponding to patterns
0-4 (single and double pixel events) selected for the pn data and patterns 0-12 for MOS. A low energy cut of 300 eV was applied to all
X-ray data and known hot or bad pixels were removed. Source counts were obtained from a circular region of 45\arcs\ radius centred on the
target source, with the background being taken from a similar region offset from, but close to, the source. Resulting exposures were 19570
s (pn) and 43149 s (combined MOS). 

The X-ray flux from \2M\ was found to have increased by a factor $\sim$10 from the earlier \xmm\ observation. Together with the longer
exposures, the brighter X-ray source meant the number of counts was greatly increased, to 11756 in the pn camera (288 in 2003) and 8034
(241) in the MOS.  As the X-ray light curve showed only small variations (but see later) the integrated pn and MOS spectra were used for
spectral fitting, each data set being grouped to a minimum of 20 counts  per bin to facilitate use of the $\chi^2$ minimalisation
technique. Spectral fitting was based on the Xspec package (Arnaud 1996), with all spectral fits including absorption due to the
line-of-sight Galactic column of $N_{H}$=4.1$\times 10^{20}$ cm$^{-2}$. Errors are quoted at the 90\% confidence level ($\Delta
\chi^{2}=2.7$ for one interesting parameter).

\begin{figure}                                                          
\centering                                                              
\includegraphics[width=4.7cm, angle=270]{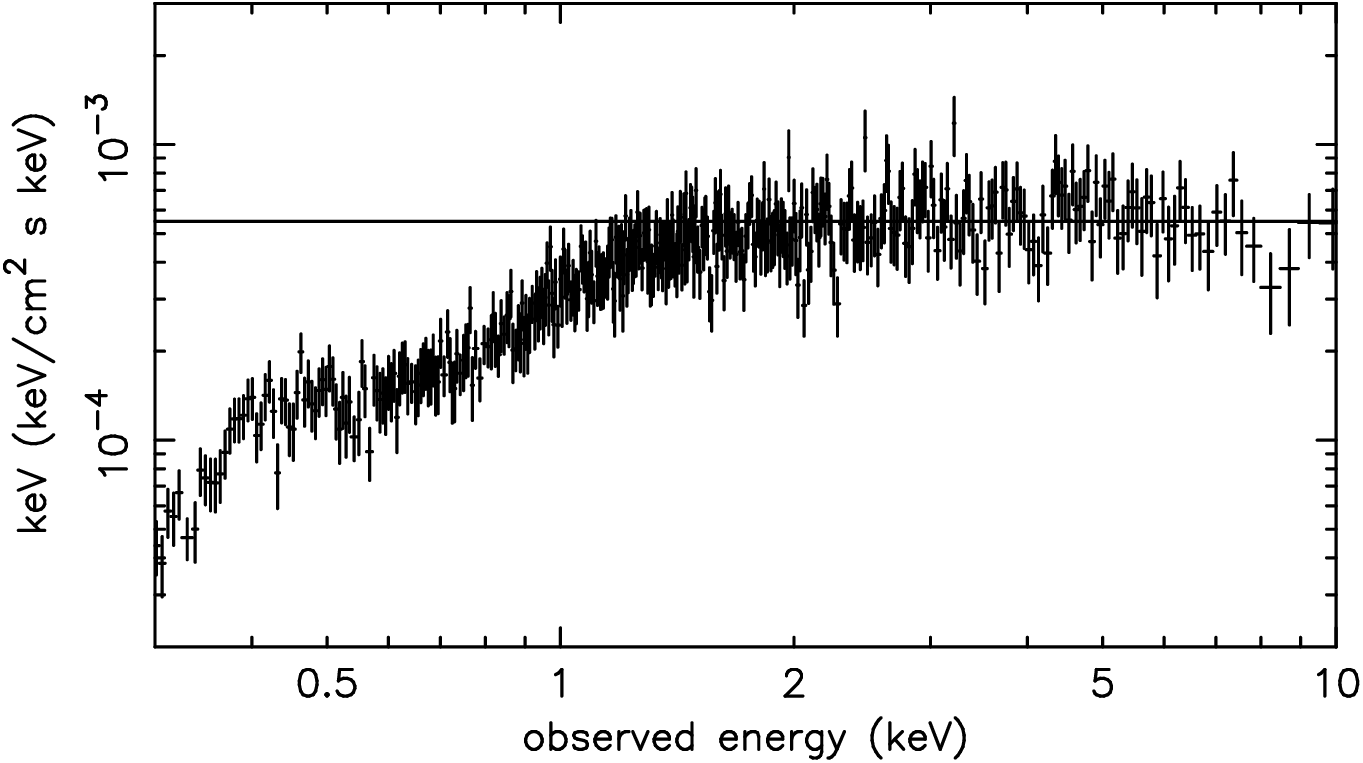}                                         
\centering                                                               
\includegraphics[width=4.7cm, angle=270]{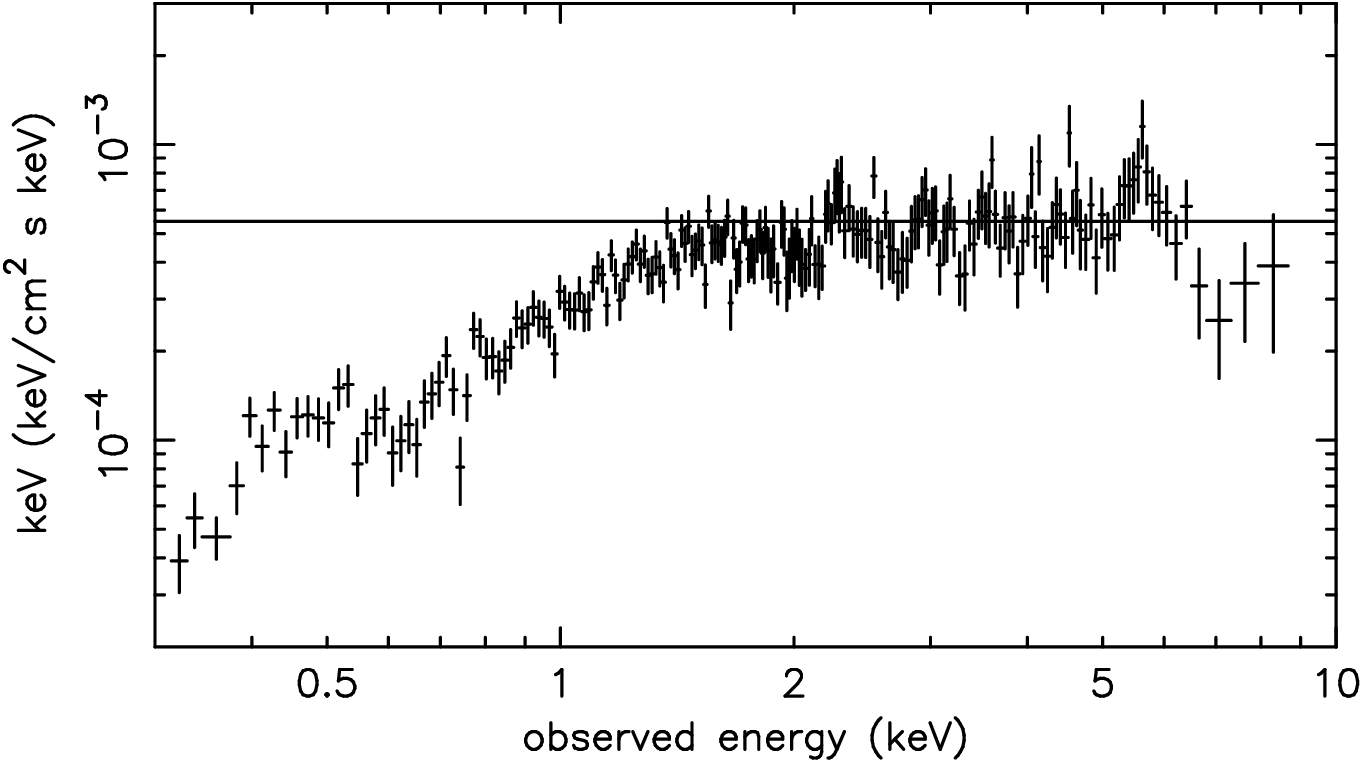}                                         
\caption                                                                
{Fluxed pn (upper panel) and MOS spectra (lower panel) from the 2005 observation of \2M. The typical spectral slope of a type 1 AGN ($\Gamma$=2 is flat in this
plot) can be  seen over the 2-6 keV band, together with clear evidence for low energy absorption}      
\end{figure}

\section{Spectral fitting}

The fluxed spectra from the 2005 observation of \2M (figure 1) show a spectral slope typical of a type 1 AGN (approximately  flat in this plot)
over the 2-6 keV  band. Below $\sim$2 keV is a clear signature of continuum absorption, while in the highest energy channels of both pn and MOS
spectra there appears to be additional absorption structure. 

Initial spectral modelling proceeded in 3 steps. First, a power law fit to the 2-6 keV data, a spectral region expected to be least affected by
absorption, was found to be statistically acceptable ($\chi^2$=220 for 201 degrees of freedom) with $\Gamma$=1.78$\pm$0.07 (pn) and
$\Gamma$=1.77$\pm$0.07 (MOS).

Extending the 2-6 keV power law fits down to 0.3 keV (figure 2a) reveals a flux deficit below $\sim$2 keV. To model this  in Xspec we added
ABSORI to the power law continuum, with free parameters of column density and ionisation parameter $\xi$ (=L/nr$^{2}$, where n is the particle
density at a distance of r from a source of ionising luminosity L). 

This gave a reasonably good fit ($\chi^2$ of 605/582), with an absorbing column of $N_{H}$$\sim$$3.7\times10^{21}$ cm$^{-2}$) of weakly ionised
gas ($\xi$$\sim$0.11 erg cm s$^{-1}$). Interestingly, inspection of the data:model ratio in this fit (figure 2b) shows some residual spectral
structure, with a hint of emission features near $\sim$0.4 and $\sim$0.6 keV, similar to those seen in the 2003 observation. To quantify those
features we then added gaussian lines to the Xspec model, finding 2  narrow ($\sigma$=35 eV) `emission lines', at $\sim$0.40 keV (0.46 keV in
the AGN rest frame) and $\sim$0.60 keV (0.69 keV) to give an excellent overall fit, with $\chi^2$  of 574/578 (figure 2c). If the soft emission
arises from a photoionised/photoexcited gas then the gaussian structures could be identified with resonance emission from He- and H-like ions of
N and O. The parameters of the best-fit absorbed power law plus gaussian line model are summarised in Table 1. 

\begin{figure}                                                          
\centering                                                              
\includegraphics[width=4.13cm, angle=270]{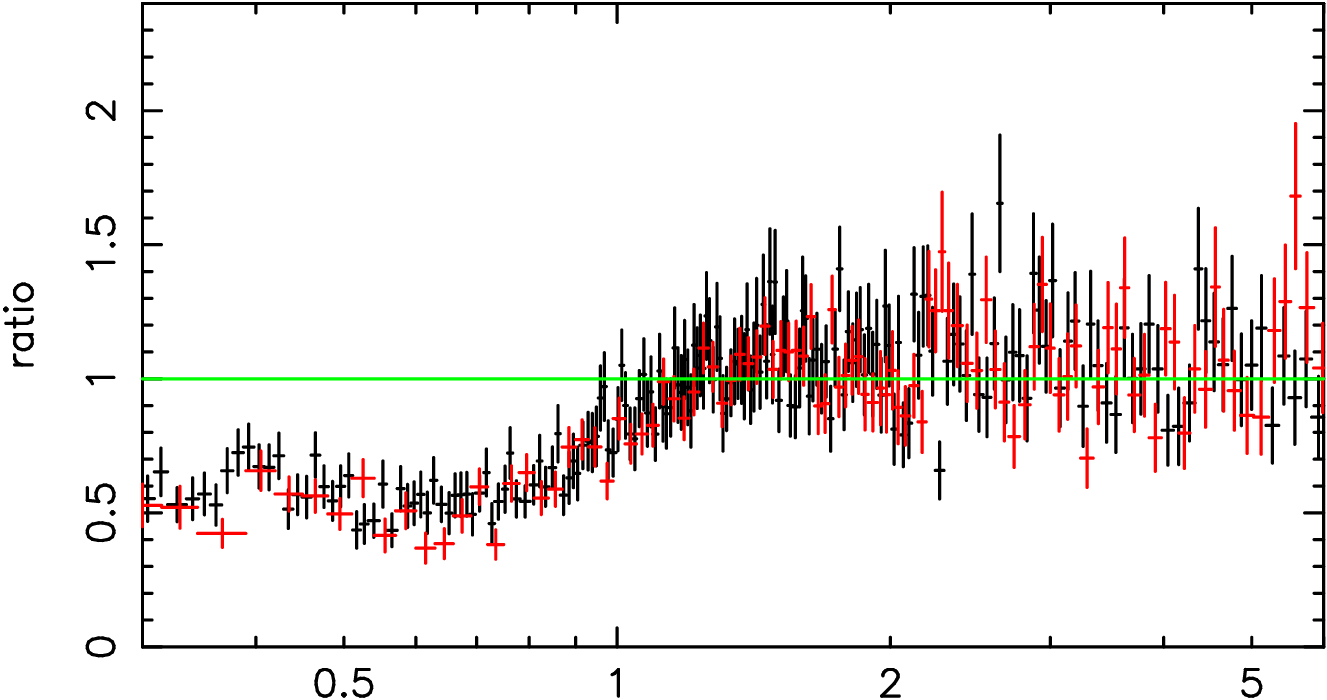}                                                                                 
\centering                                                              
\includegraphics[width=4.13cm, angle=270]{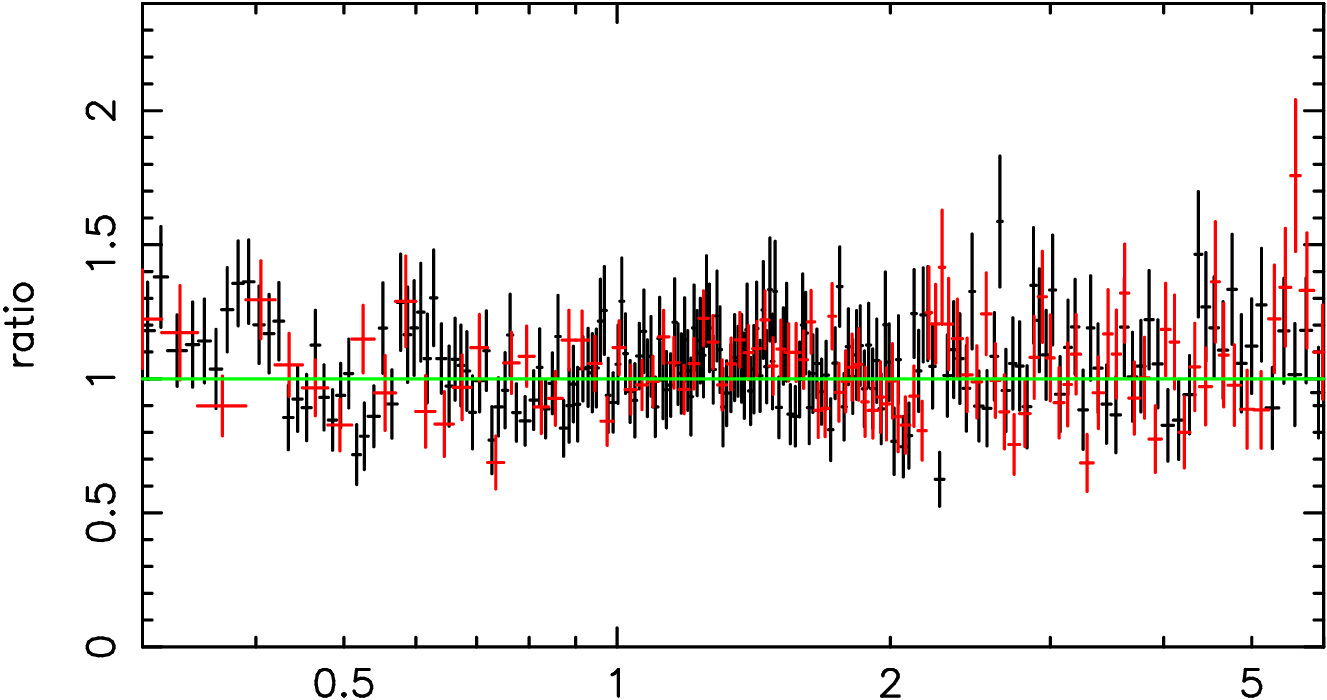}                                         
\centering                                                               
\includegraphics[width=4.5cm, angle=270]{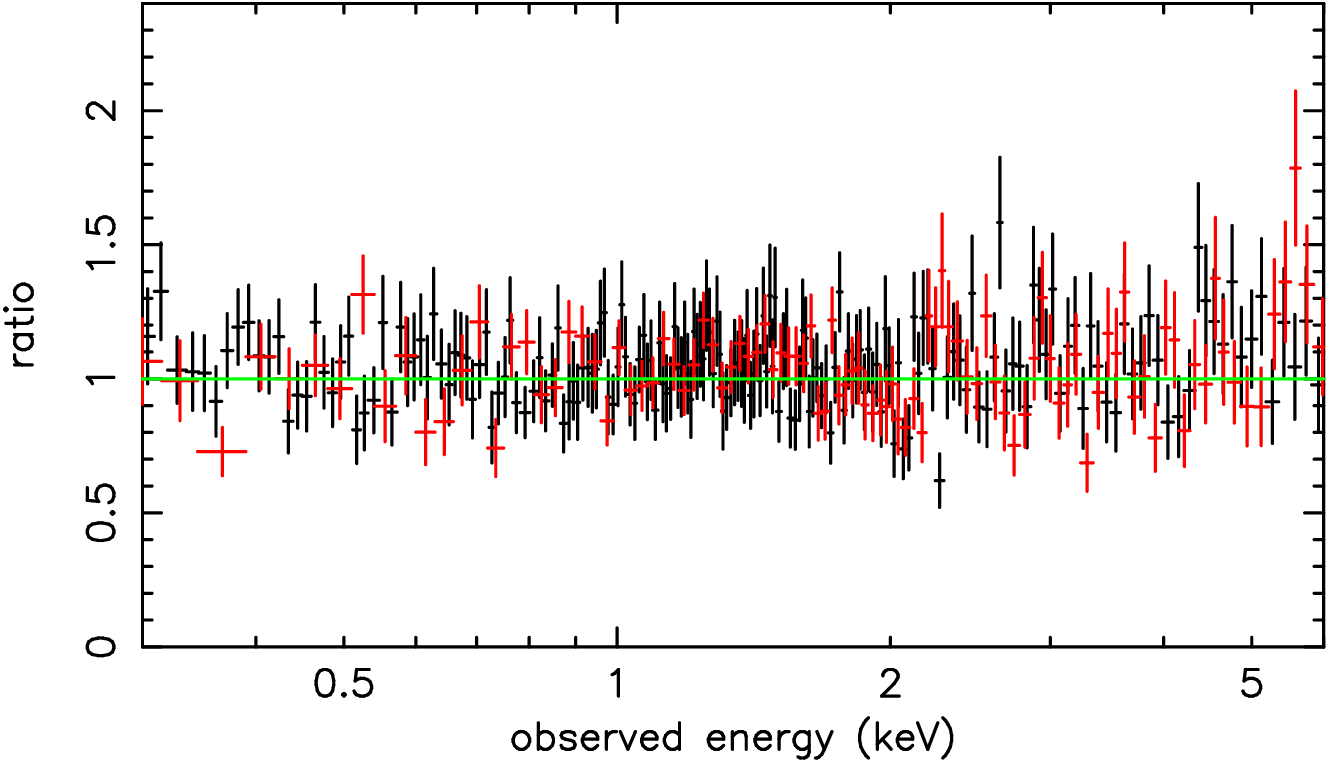}                                         

\caption                                                                
{From the top are shown the pn (black) and MOS (red) data:model ratios over the 0.3-6 keV band for a) the 2-6 keV power law continuum, b) power law plus weakly ionised
absorber, and c) with the further addition of gaussian line emission, as described in the text}      
\end{figure}                                                            

Based on the above broad band fit, the X-ray luminosity of \2M\ can be computed, finding over the
full 0.3-10 keV band an observed luminosity of 1.2$\times 10^{44}$ erg s$^{-1}$, a factor $\sim$8.5 greater than in
the 2003 observation. In the soft X-ray band (0.3-1 keV) the luminosity was 1.4$\times 10^{43}$ erg s$^{-1}$, with $\sim$$2.7\times
10^{42}$ erg s$^{-1}$ in the soft emission component. 

The absorption-corrected 2-6 keV luminosity of 8.5$\times 10^{43}$ erg s$^{-1}$ allows an estimate to be made of the bolometric
luminosity of \2M, using the scaling established for a wide range of AGN (Marconi \et\ 2004). We find L$_{bol}$$\sim$2.5$\times
10^{45}$ erg s$^{-1}$.

\begin{table*}
\centering
\caption{Summary of the fit parameters for \2M\ over the 0.3-10 keV band. As described in the text the model is a power law, with weakly ionised
absorption plus residual soft X-ray emission represented by gaussian lines g1 and g2. Line energies are in the source rest frame and fluxes are in 
units of $10^{-5}$ph cm$^{-2}$ s$^{-1}$}

\begin{tabular}{@{}lccccccc@{}}
\hline
camera & $Gamma$ & $N_{H}$ (cm$^{-2}$) & $\xi$(ergs cm s$^{-1}$) &  g1 (keV) & g1 flux &  g2 (keV) & g2 flux \\

\hline
pn & 2.12$\pm$0.11 & 4.0$\pm$0.3$\times 10^{21}$ & 0.03$\pm$0.02 & 0.69$\pm$0.02 & 2$\pm$1.5 & 0.46$\pm$0.02 & 8$\pm$4 \\
MOS & 2.04$\pm$0.11 & 4.0$\pm$0.3$\times 10^{21}$ & 0.03$\pm$0.02 & 0.69$\pm$0.03 & 2$\pm$1.5 & 0.46$\pm$0.02 & 8$\pm$4 \\

\hline
\end{tabular}
\end{table*}

\section{The nature of the spectral change between the low and high flux states} 
The spectral form observed in the low and high flux \xmm\ observations of \2M\ are markedly different, with the \chandra\ spectrum
sitting closer in form and flux level to the latter. The high flux\xmm\ spectrum is the best defined and can be understood in terms
of a typical type 1 AGN X-ray continuum attenuated at low energies by an absorbing column which - if `cold' - may also explain the
strong nuclear reddening of \2M. Weak residual soft X-ray features are consistent with the soft excess seen in the
low flux spectrum.

Least well defined, as a result of the much lower counts, is the nature of the hard X-ray spectrum in the 2003 \xmm\ observation. W05
suggested this was due to strong reflection based on an upturn in the highest energy channels. Importantly, given the optical
characteristics of \2M, absorption could not be usefully constrained in that observation.  Having now found a substantial cold
column in the 2005 spectrum, could a much larger column have caused the hard spectral form in 2003?

The ratio of spectral data between observations at different flux levels is an established way of testing for variable
multiplicative components, such as absorption. To check the above suggestion we therefore calculated the spectral ratios of the low
to high flux observations of \2M\ for both pn and MOS data. The results were similar and are reproduced for the pn data in figure 3. 
The low-to-high flux ratio is flat
over the $\sim$1-5 keV band, but increases at both low and high energy ends of the spectrum. Although the statistics are rather
poor, limited by the low flux data points, the flat mid section of the ratio plot {\it does not} have the form of variable absorption
(in contrast, for example, with 1H0419-577; Pounds \et\ 2004). 

On the other hand, the increases in the lowest and highest energy channels {\it are} consistent with the cold reflection and soft
emission components suggested in the low flux modelling (W05) being less variable than the power law continuum.
A similar finding has been reported from extended studies of the bright type 1 Seyferts MCG-6-30-15 (Fabian and Vaughan 2003) and
1H 0419-577 (Pounds \et\ 2004).

\begin{figure}                                                          
\centering                                                              
\includegraphics[width=4.7cm, angle=270]{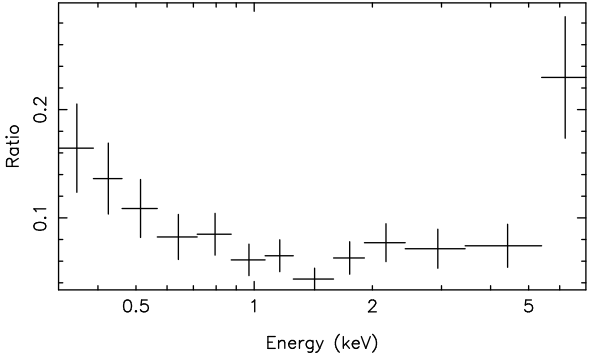}                                                                                                 
\caption                                                                
{Ratio of low to high flux data for the pn camera, supporting the indication from spectral modelling that cold
reflection  and a soft emission  component are less evident when \2M\ is in a bright state. Importantly, the ratio plot is not consistent with
the main spectral change being due to an increase in low energy absorption  in the  low flux state}      
\end{figure}                                                            

The above interpretation of the ratio plots, where the dominant spectral change is driven by a variable strength type 1 power law, is
supported by visual examination of the data sets. This is illustrated in figure 4 where the pn data from both high and low flux
observations is compared with a simple power law continuum, fitted with a common power law index ($\Gamma$$\sim$1.8) over the
restricted 2-5 keV band. The power law is seen to match {\it both} data sets quite well over an intermediate spectral band where absorption and
reflection are least likely to  be seen.

\begin{figure}                                                          
\centering                                                              
\includegraphics[width=5.8cm, angle=270]{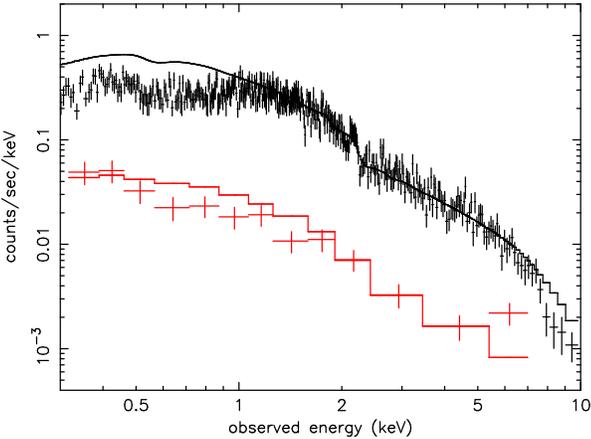}                     
\caption                                                                
{Comparison of data and model after fitting both 2003 (red) and 2005 (black) pn camera spectra to a common power law index between 2-5 keV 
($\Gamma$=1.8)}      
\end{figure}                                                            

\subsection{Absorption and emission features in the Fe-K band}
The fluxed spectra in figure 1 show evidence for absorption above $\sim$6 keV in both pn and MOS data, while the MOS spectrum also shows a possible
emission line close to the energy of neutral Fe-K. When re-plotted as a ratio of data to the 2-6 keV power law fit (figure 5) the pn and MOS features
appear different, particularly in the more obvious Fe-K emission line and lower energy of the absorption `line' in the MOS spectrum. Although the MOS
sensitivity falls more steeply than the pn at high energies,  in-flight data show the absolute energy calibration of the pn and MOS cameras to agree
to within a few eV in the Fe K band. However, there are well-known - and different - features in the EPIC background spectra (Str\"{u}der \et\ 2001,
Turner \et\ 2001). To explore whether
such features in the background could be contributing  to the apparent high energy spectral stucture we show, in figure 6, directly comparable source
and background spectra, obtained from regions of the same area on each \xmm\ image. 

These plots raise doubts about the background-subtracted spectra above $\sim$7 keV in the MOS and above $\sim$8 keV in the pn data.
We therefore decided to assess spectral structure in the Fe-K band by fitting the source data, without background subtraction, for the pn 
and MOS cameras combined. Starting with the extrapolated 2-6 keV power law fits, we sequentially added an emission line and 2
absorption edges/lines to the power law continuum. 
A narrow emission line at 5.59$\pm$0.07 keV (rest frame 6.42$\pm$0.07 keV) was only marginally significant in the combined
data set, with an EW of 75$\pm$50 eV and a small improvement in the fit from $\chi^2$ of 269/228 to $\chi^2$ of 264/226. An absorption line at 
6.88$\pm$0.05 keV (rest frame 7.91$\pm$0.05 keV), of width $\sigma$$\sim$100 eV and EW 360$\pm$50 eV, produced a larger improvement, to $\chi^2$ of 
236/223. Finally, an absorption edge at 7.6$\pm$0.2 keV (rest frame 8.7$\pm$0.2 keV), optical depth $\tau$ of 0.3$\pm$0.1, further improved the fit to $\chi^2$ of 228/221.

We defer further discussion of these features to Section 5.3.

\begin{figure}                                                          
\centering                                                              
\includegraphics[width=5.8cm, angle=270]{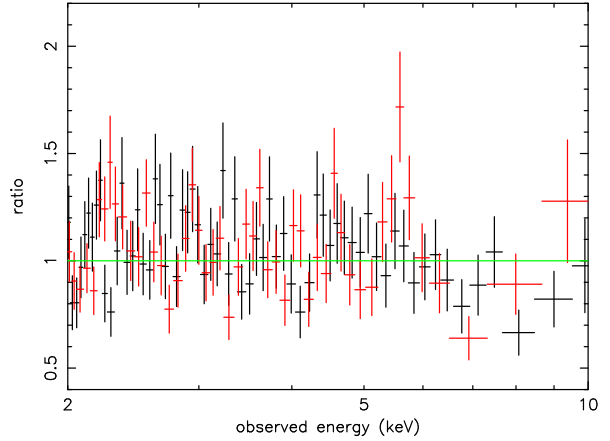}                     
\caption                                                                
{Ratio of source-only data to the 2-6 keV power law fit for the 2005 observation of \2M. The pn (black) and MOS (red) source spectra show 
differences in apparent
emission and absorption features in the Fe-K band}      
\end{figure}                                                            

\begin{figure}                                                          
\centering                                                              
\includegraphics[width=4.7cm, angle=270]{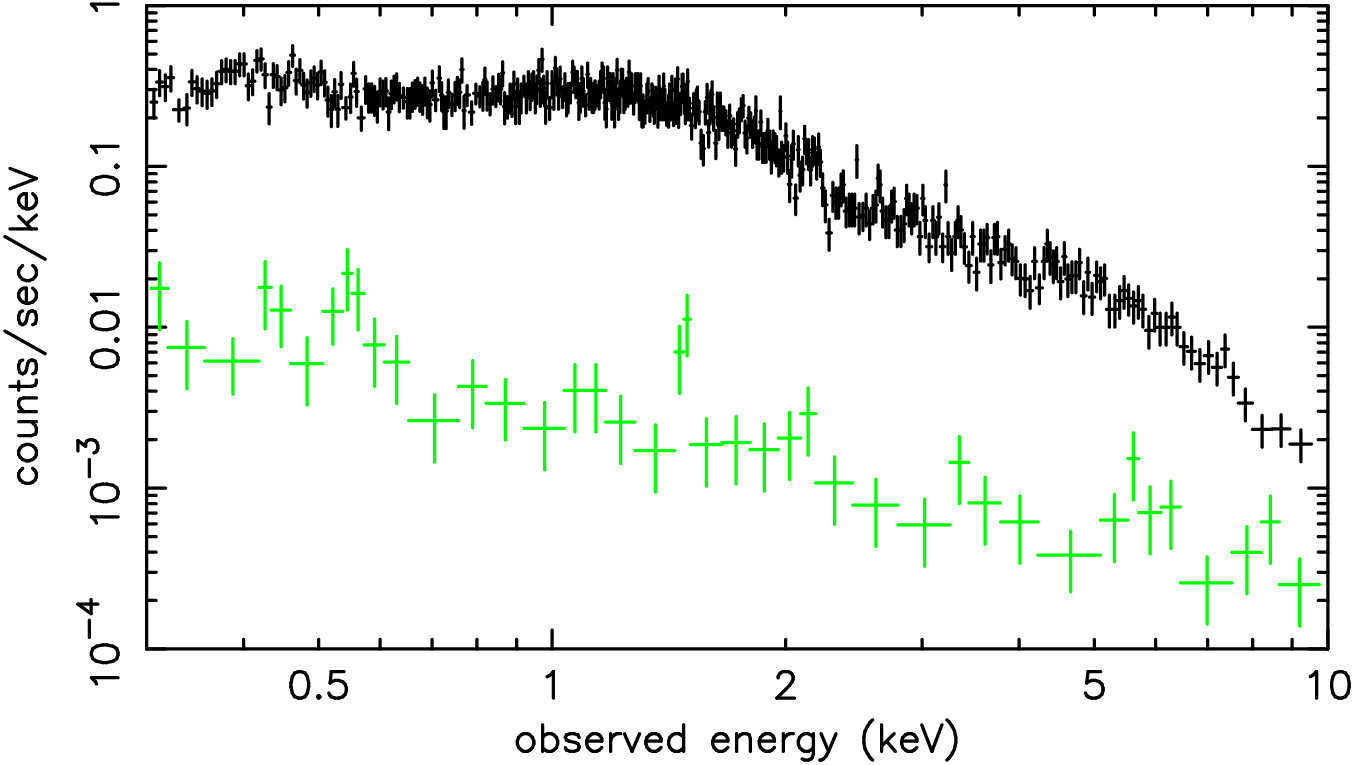}                     
\centering                                                              
\includegraphics[width=4.7cm, angle=270]{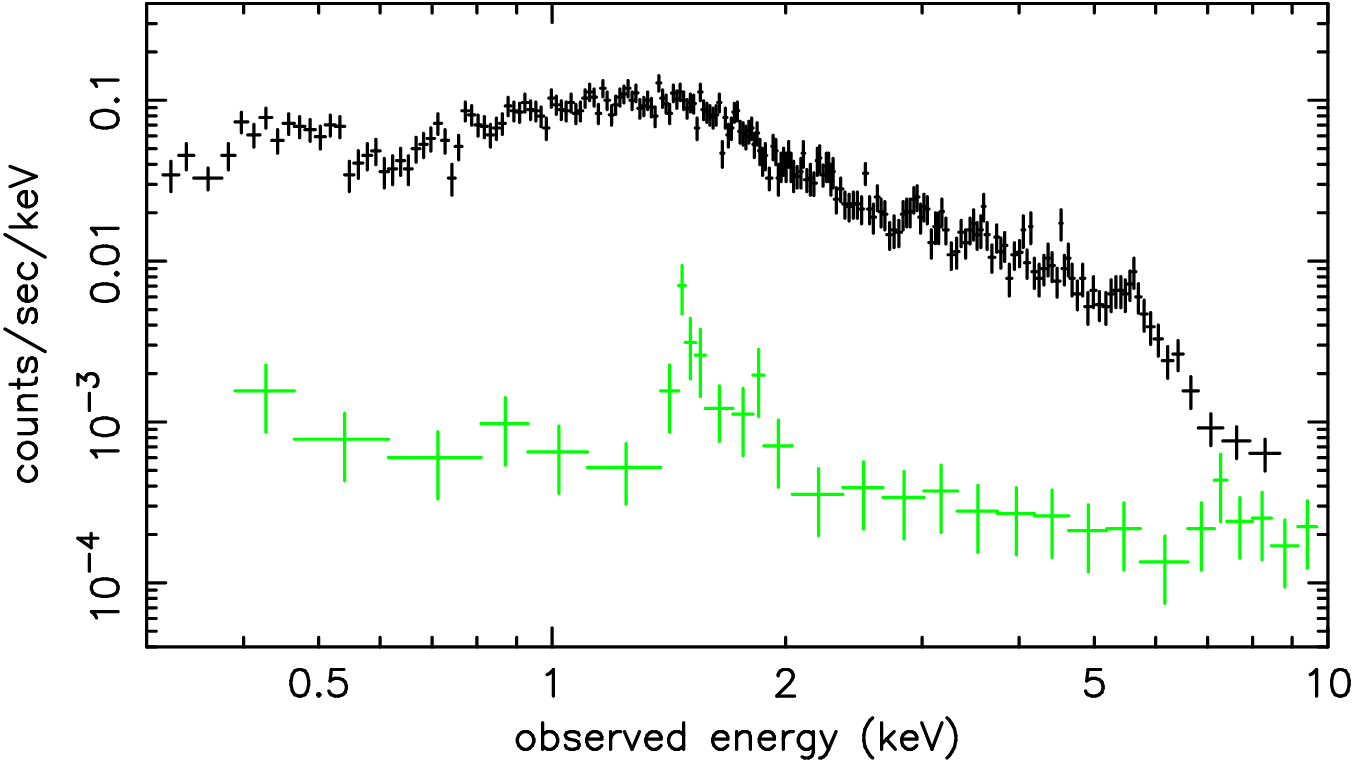}                     
\caption                                                                
{Source and background spectra showing the increases in background count rate near $\sim$6 keV and$\sim$8 keV (pn camera, upper panel) 
and $\sim$7 keV (MOS camera, lower panel) which raise doubts about the apparent high energy absorption features in figures 1 and 5}      
\end{figure}

\section{Discussion}
\subsection{Low energy absorption}
The marginal detection of low energy absorption in the original \chandra\ observation of \2M (Wilkes \et\ 2002) was a surprise, given that
\2M\ is a highly reddened  object.  On the basis of the follow-up \xmm\ observations it now seems that the
intermediate flux level \chandra\ observation may have been confused by the presence of a soft emission component, seen more clearly
in the low flux state \xmm\ observation in 2003. We now find, in the 2005 high state spectrum, clear evidence for low energy
absorption corresponding to a line-of-sight column N$_{H}$$\sim$4$\times 10^{21}$  cm$^{-2}$. With the low ionisation
parameter and a `normal' dust to gas ratio, that absorbing column is consistent with the additional optical reddening 
(i.e. above that of an unobscured QSO; Barkhouse and Hall 2001) of J-K$_{s}$$\sim$0.4 seen in \2M.

Our analysis suggests that the spectral change between the 2003 and 2005 \xmm\ observations was dominated by a change in the X-ray 
power law continuum, with absorption in a column of
N$_{H}$$\sim$4$\times 10^{21}$  cm$^{-2}$ of  `cold' or weakly ionised gas present in both cases. As noted in W05,
inclusion of such absorption in modelling the 2003 \xmm\ spectrum, in addition to strong cold reflection, would allow the underlying 
continuum to assume a `normal' photon
index, in the range predicted by Comptonisation models (eg Haardt \et\ 1997). Variability of the primary power law continuum has been identified as 
the dominant component in X-ray spectral changes
in several well studied AGN (Fabian and Vaughan 2003, Pounds \et\ 2004). It appears that \2M\ is a further example of that situation.

While we have no direct information on the location of the absorbing gas, it is probably not very close to the central source unless
in sufficiently dense clouds to avoid evaporation. However, the intermediate type 1.5 classification of \2M\ suggests some absorbing matter is located
at a radius comparable
with the BLR. Future observations to check for variability of the absorbing column and/or ionisation parameter should clarify that issue.

\subsection{Soft X-ray emission}
It is interesting to speculate on the nature of the soft X-ray emission which, although  obviously not well constrained in either
case, is seen in both the 2003 and 2005 observations. At face value the luminosity of the structured soft X-ray emission is higher
in the 2005 observation, with $2.7\times 10^{42}$ergs s$^{-1}$ in the 2 gaussian components, a factor $\sim$4 greater than the
similarly defined `soft excess' in the 2003 observation. While the `soft excess' is not determined independently of either the absorbing column or
the strength and slope of the power law continuum, light curves for 0.3-1.0 keV and 1-10 keV data (figure 7) indicate that the soft
flux is less variable within the 2005 observation, suggesting a significant soft component is separate from (and more extended than) the power law
source. 

To obtain a further measure of the longer-term variability of the soft X-ray component we have tried fitting the 2003 data to the 2005 spectral model
(absorbed power law plus soft excess) with the power law slope and cold absorber fixed. The major difference was found to be in the normalisation of
the power law, which fell by a factor of $\sim$14. Half the remaining excess in chi$^{2}$ was removed by allowing the power law index to change, a
reduction from $\sim$2.1 to $\sim$1.9 perhaps corresponding to the relatively stronger reflection in the low flux spectrum. Finally, the soft X-ray 
emission was allowed to vary, an acceptable fit being obtained for a fall in gaussian line flux of a factor $\sim$1.6.

\begin{figure}                                                          
\centering                                                              
\includegraphics[width=5.8cm, angle=270]{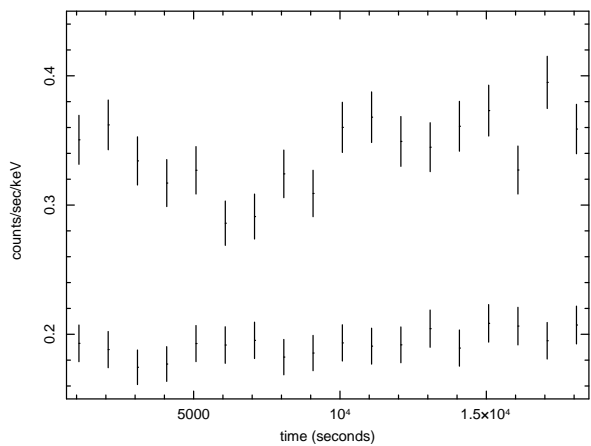}                     
\caption                                                                
{X-ray light curve from the 2005 observation. The upper plot for the 1-10 keV band shows evidence for variability on the timescale of 2-3000 seconds
which is not replicated in the soft 0.3-1 keV flux}      
\end{figure}

While the above estimate is model dependent, and a constant soft flux is not ruled out, it is clear that the 
soft X-ray emission component, significant in both \xmm\ observations, is much less variable than the power law continuum.
The relative lack of variability supports an origin of the soft X-ray emission in an extended outflow, as is actually resolved in nearby type 2
AGN. 
That physical link has been proposed in the analysis of the X-ray spectra of other type 1 AGN and a quantitative comparison made in Pounds \et (2005). 
We revise this comparison here and add our new analysis for \2M.

Table 2 summarises the results, where the 2-10 keV luminosities, corrected for absorption in our line of sight, are used as a proxy for the ionising flux 
irradiating the soft X-ray 
emission region. The data indicate a trend where the
relative strength of the soft X-ray emission in type 1 AGN is typically an order of magnitude greater than for type 2 AGN. In that context the
intermediate value for \2M\ is consistent with its intermediate optical type, supporting the view that the soft X-ray emission typically arises
from an ionised outflow originating at a smaller radius the BLR, possibly driven off the same cold matter seen in absorption. In that context we note that 
energetically the soft X-ray emission in the 2005 observation of \2M\ corresponds to only $\sim$5 \% of the power law continuum flux removed by the cold absorber.

\begin{table*}
\centering
\caption{The soft X-ray and 2-10 keV luminosities of \2M\ from the 2005 observation, together with comparable luminosities 
for 3 Type 1 AGN and for the archetypal Type 2 AGN, Mkn3 and NGC 1068. The soft X-ray emission luminosities are as observed and as
corrected for absorption in our Galaxy.  With the 2-10 keV luminosities, corrected for intrinsic absorption, used as a proxy for the
ionising flux irradiating the soft X-ray  emission region in each case, the table suggests a trend with a major fraction of the
ionised outflow being hidden from our view in type 2 objects. This is an expanded and revised version of Table 1 in Pounds \et (2005)}

\begin{tabular}{@{}lccccccc@{}}
\hline
Galaxy & optical type & L$_{2-10}$(ergs s$^{-1}$) & L$_{SX}$(ergs s$^{-1}$) &  ratio (L$_{SX}$/L$_{2-10}$) & L$_{SX}$(ergs s$^{-1}$) 
&  ratio (L$_{SX}$/L$_{2-10}$) \\
\hline
NGC 4051 & Seyfert 1 & $3.5\times 10^{41}$ & $3.5\times 10^{40}$ & 10\% & $3.5\times 10^{40}$ & 10\% \\
1H 0419-577 & QSO 1 & $2.5\times 10^{44}$ & $3\times 10^{43}$ & 12\% & $3.2\times 10^{43}$ & 13\% \\
2M 2344+1221 & QSO 1 & $5.2\times 10^{43}$ & $6.1\times 10^{42}$ & 11\% & $10.1\times 10^{42}$ & 19\% \\
Mkn 3 & Seyfert 2 & $1.5\times 10^{43}$ & $1.6\times 10^{41}$ & 1.1\%  & $2.4\times 10^{41}$ & 1.6\% \\
NGC 1068 & Seyfert 2 & $2.3\times 10^{43}$ & $1.5\times 10^{41}$ & 0.6\%  & $1.7\times 10^{41}$ & 0.7\% \\
2M 0918+2117 & QSO 1.5 & $8.6\times 10^{43}$ & $2.7\times 10^{42}$ & 3.1\%  & $3.8\times 10^{42}$ & 4.4\% \\
\hline
\end{tabular}
\end{table*}


\subsection{Evidence for absorption in highly ionised gas?}

Analysing the high energy spectrum of \2M\ is unusually challenging due to the presence of apparently quite strong emission and absorption features 
which, however,
are not consistent between the pn and MOS data. These differences remain, although at a reduced level, when the background subtraction is removed.
Finding no other explanation than limited statistics we therefore analysed the source-only spectra for the pn and MOS data combined.

We find a marginally significant emission line at an energy consistent with Fe-K fluorescence from low ionisation matter. The equivalent width is consistent
with continuum reflection being much less significant (relative to the direct power law) than in the 2003 observation. At the low level detected, scattering from the cold absorber might be a
candidate source.

A stronger fall in sensitivity in the Fe K band would explain why the MOS camera only detects the absorption feature observed at $\sim$6.9 keV ($\sim$7.9 keV 
in the AGN rest frame).
Figure 5 suggests an absorption line, rather than an absorption edge, and indeed that choice is statistically preferred. If real, the most likely
identification is with resonance line absorption in highly ionised Fe. For He-like FeXXV the measured line energy would imply an outflow velocity of
v$\sim$0.15c.
  
The broader absorption feature near $\sim$7.6 keV ($\sim$ 8.7 keV in the AGN rest frame), only detected in the pn camera, is better modelled with an
absorption edge. Intriguingly, the K-edge of He-like FeXXV (threshold energy 8.83 keV; Verner \et\ 1996) lies close to the derived edge energy. With that
identification and a threshold cross section of 2$\times 10^{-20}$ cm$^{2}$ (Verner \et\ 1996) the measured edge optical depth of $\sim$0.3 implies
an absorbing column density of N$_{FeXXV}$$\sim$1.5$\times 10^{19}$ cm$^{-2}$. Assuming Fe XXV to be the dominant ion (likely over a rather wide range
of ionisation parameter in a highly ionised gas, and noting no evidence for an FeXXVI edge),  this corresponds, for a solar abundance of Fe, to a column density of
N$_{H}$$\sim$5$\times 10^{23}$ cm$^{-2}$.

There have been an increasing number of reports of highly ionised gas imprinting Fe K features on AGN X-ray spectra. Due to the relatively low cross
sections at those energies, the derived column densities are always high. One of the first (Hasinger \et\ 2002) suggested an interpretation of
features in the high redshift BAL quasar APM08279+5255 as FeXV-FeXVIII edge absorption in a column density of  N$_{H}$$\sim$$10^{23}$  cm$^{-2}$.
More often, higher quality spectra have shown absorption lines of FeXXV or FeXXVI, in some cases also indicating rapidly outflowing gas (Pounds \et\
2003,2006, Reeves \et\ 2003, Young \et\ 2005)).

But are the absorption features seen in the present spectra of \2M\ real? Visual examination of the respective background spectra shows the
background rate rising strongly above $\sim$6 keV in the MOS and  $\sim$7 keV in the pn data. While the pn camera background is also enhanced near
$\sim$8 keV (probably due to the Cu K line arising from the electronics board, Str\"{u}der \et\ 2001), the `absorption edge' observed at $\sim$7.6 keV  is 
statistically
significant in the source-only data. 

In conclusion, while continuing to regard the absorption features in the Fe K band as doubtful, we believe they are of sufficient  potential interest
to retain for future checking with better data.

\section{Summary}

1)Analysis of a new \xmm\ observation of the type 1.5 QSO \2M\ confirms the presence of a cold
absorbing column consistent with the red optical nucleus and intermediate optical type. 

2)Comparison with the previous \xmm\ observation when \2M\ was in a much lower flux state, provides further evidence that strong variability in AGN X-ray spectra 
arises when a relatively steep ($\Gamma$$\ga$2) power law continuum overcomes a more
complex and quasi-constant underlying spectrum. Previous explanations of hard low state spectra ($\ga$2 keV) being reflection
dominated appear also to apply to \2M.  

3)Spectral structure seen below $\sim$1 keV in the 2003 \xmm\ observation of \2M\ and interpreted as emission from ionised gas, appears again to be 
present in the 2005 observation.

4)Comparison of the relative strength of the soft X-ray emission in a number of type 1 and type 2 AGN shows the former to be typically
an order of magnitude greater, suggesting ionised outflows originate well within the BLR and so are substantially obscured in type 2 sources. 
The intermediate value found here
for \2M\ may then be consistent with its intermediate optical type.

5)Evidence for strong absorption in the higher energy channels of both EPIC cameras is reduced but is still statistically significant when background features 
are removed. If real, the high energy absorption implies
a large column of highly ionised gas in the line of sight to \2M.  

\section{Acknowledgments} 
The results reported here are based on observations obtained with \xmm, an ESA science
mission with instruments and contributions directly funded by ESA Member States and the USA (NASA). The authors wish
to thank the SOC and SSC teams for organising the \xmm\ observations  and initial data reduction.  BJW is grateful for the financial support 
of \xmm\ GO grant:
NNG04GD27G.


\begin{thebibliography}{}
\bibitem{} Arnaud K.A.  \ 1996, ASP Conf. Series, 101, 17 
\bibitem{} Barkhouse W.A., Hall P.B. \ AJ, 121, 2843 
\bibitem{} Cutri R., \et, \ 2001, "New Era of Wide Field Astronomy", Ed. R.Clowes, A.Adamson, G. Bromage, 
ASP Conf. Ser., 32, 78 
\bibitem{} Cutri R., \et, \ 2003, on-line catalogue II/246, IPAC/Cal Tech
\bibitem{} Fabian A.C., Vaughan S. \ 2003, MNRAS, 340, L28
\bibitem{} Haardt F., Maraschi L., Ghisellini G. \ 1997, ApJ, 476, 620 
\bibitem{} Hasinger G., Schartel N., Komossa S. \ 2002, ApJ, 573, L77
\bibitem{} Marconi A., Risaliti G., Gilli R., Hunt L.K., Maiolino R., Salvati M. \ 2004, MNRAS, 351, 169
\bibitem{} Nandra K., Pounds K.A. \ 1994, MNRAS, 268, 405
\bibitem{} Pounds K.A., Reeves J.N., King A.R., Page K.L., O'Brien P.T., Turner M.J.L \ 2003, MNRAS, 345, 705
\bibitem{} Pounds K.A., Reeves J.N., Page K.L., O'Brien P.T. \ 2004, ApJ, 616, 696
\bibitem{} Pounds K.A., Wilkes B.J., Page K.L. \ 2005, MNRAS, 362, 784
\bibitem{} Pounds K.A., Page K.L. \ 2006, MNRAS, 372, 1275
\bibitem{} Reeves J.N., O'Brien P.T., Ward M.J. \ 2003, ApJ, 593, L65
\bibitem{} Smith P.S., Schmidt G.D., Hines D.C., Cutri R.M., Nelson B.O. \ 2002, ApJ, 569, 23  
\bibitem{} Smith P.S., Schmidt G.D., Hines D.C., Foltz C.B. \ 2003, ApJ, 593, 676  
\bibitem{} Str\"{u}der L.\et, \ 2001, A\&A, 365, L18 
\bibitem{} Turner M.J.L. \et, \ 2001, A\&A, 365, L27 
\bibitem{} Verner D.A., Ferland G.J., Korista K.T., Yakovlev D.G. \ 1996, ApJ, 465, 487
\bibitem{} Wilkes B.J., Schmidt G.D., Cutri R.M., Ghosh H., Hines D.C., Nelson B. Smith P.S. \ 2002, ApJL, 564, L65
\bibitem{} Wilkes B.J., Pounds K.A., Schmidt G.D., Smith P.S., Cutri R.M., Ghosh H., Nelson B., Hines D.C. \ 2005, ApJ, 634, 183 (W05)
\bibitem{} Young A.J, Lee J.C., Fabian A.C., Reynolds C.S., Gibson R.R., Canizares C.R. \ 2005, ApJ, 631, 733



\end{thebibliography}
\end{document}